\documentclass[conference]{IEEEtran}
\usepackage{xcolor}
\definecolor{delim}{RGB}{20,105,176}
\definecolor{punct}{rgb}{0.44, 0.44, 0.44}
\definecolor{numb}{rgb}{0.1, 0.4, 0.8}
\usepackage{hyperref}

\usepackage{amsmath,amssymb,amsfonts}
\usepackage{graphicx}
\usepackage{tcolorbox}
\usepackage{textcomp}
\usepackage{enumitem}
\usepackage{listings}
\usepackage{siunitx}
\usepackage{subfigure}
\usepackage{caption,multirow}
\usepackage{makecell}
\usepackage{bbold}
\usepackage{geometry}    
\usepackage{comment}
\usepackage{threeparttable}
\usepackage{float}
\usepackage[super]{nth}
\usepackage[colorinlistoftodos]{todonotes}

\newcommand{\sixptsize}{\fontsize{6.5pt}{5.8pt}\selectfont}
\usepackage[utf8]{inputenc}
\usepackage[T1]{fontenc}

\usepackage{tikz}
\usetikzlibrary{arrows.meta, positioning, shapes.geometric}

\usepackage{algorithmic}
\usepackage[ruled,vlined,linesnumbered]{algorithm2e}

\usepackage{booktabs}     

\lstdefinelanguage{json}{
  basicstyle=\ttfamily\small,
  numberstyle=\tiny,
  stepnumber=1,
  numbersep=5pt,
  showstringspaces=false,
  breaklines=true,
  frame=single,
  backgroundcolor=\color{gray!5},
literate=
    {:}{{{\color{punct}{:}}}}{1}
    {,}{{{\color{punct}{,}}}}{1}
    {\{}{{{\color{delim}{\{}}}}{1}
    {\}}{{{\color{delim}{\}}}}}{1}
    {[}{{{\color{delim}{[}}}}{1}
    {]}{{{\color{delim}{]}}}}{1},
}

\def\BibTeX{{\rm B\kern-.05em{\sc i\kern-.025em b}\kern-.08em
    T\kern-.1667em\lower.7ex\hbox{E}\kern-.125emX}}

\begin{document}

\title{
An LLM-Based Framework for Intent-Driven Network Topology Design 
\\
}

\author{
\IEEEauthorblockN{Kholoud El Habbouli, Fen Zhou and Stéphane Huet}
\IEEEauthorblockA{\textit{† CERI-LIA,  University of Avignon, France}\\ Emails :
\{firstname.lastname@univ-avignon.fr\}}
}

\maketitle

\begin{abstract}
Designing deployable and resilient network topologies from natural language requirements remains a challenging problem in network automation. This work investigates the ability of Large Language Models (LLMs) to generate structurally valid and constraint-compliant network topologies through a constraint-driven pipeline combining hierarchical modeling and systematic validation. The framework is evaluated via a multimodel comparison of proprietary and open-weight LLMs across four realistic network scenarios released as a public dataset. We assess structural correctness using node and edge F1-scores against reference topologies, and evaluate resilience through server and content connectivity metrics. In addition, we analyze common failure modes, including interface mismatches and directional inconsistencies in generated topologies. Overall, this work provides a systematic benchmark for understanding how LLMs handle structural and resilience constraints in topology synthesis, and supports informed model selection for AI-driven network design.

\end{abstract}


\begin{IEEEkeywords}
Intent-Based Networking, Network Resilience, Large Language Model (LLM), Topology Synthesis, Structural Fidelity.
\end{IEEEkeywords}

\section{Introduction}
Recent large-scale outages in cloud and 5G infrastructures have exposed a critical limitation in current network automation systems: correct configuration alone does not guarantee operational resilience. As network scale and complexity increase, ensuring reliable service delivery requires reasoning not only at the configuration level, but also at the level of network topology design.

Existing approaches in Intent-Based Networking (IBN) \cite{10664144} and Large Language Model (LLM)-driven automation primarily focus on translating high-level intents into device configurations, assuming predefined and valid network topologies. Frameworks such as \textit{NetConfEval} \cite{netconfeval} and \textit{S-Witch} \cite{jeong2024switch} treat topology as an input rather than a design output, which limits their applicability in scenarios requiring explicit structural constraints and redundancy guarantees. As a result, LLM-based systems may generate structurally inconsistent or invalid network representations when reasoning about connectivity and interface compatibility.

To address this limitation, we argue that resilience must be considered at the topology design stage, prior to configuration synthesis. We introduce a Resilience-by-Design perspective, where network structures are generated under explicit constraints ensuring consistency and redundancy before deployment. This decouples topology synthesis from configuration generation, enabling a structured and verifiable design process. A key challenge in this domain is the lack of benchmarks for evaluating LLM-based network topology synthesis under realistic structural and resilience constraints. To address this gap, we introduce a benchmark composed of intent-to-topology scenarios, together with an evaluation methodology for systematically assessing and comparing LLM-generated network topologies. 

This paper makes the following contributions:
\begin{itemize}
\item To the best of our knowledge, we are the first to propose an LLM-based framework for intent-driven network topology design that comprehensively incorporates network structure, heterogeneous network devices and interface diversity, as well as resilience requirements.
\item We propose a benchmark for intent-driven topology synthesis under realistic scenarios.
\item We design an evaluation methodology combining node and edge F1-scores with connectivity-based resilience metrics.
\end{itemize}

\begin{figure*}[!h]
\centering
\subfigure[Designed topology (Scenario~1).]{
\includegraphics[width=0.3\textwidth]{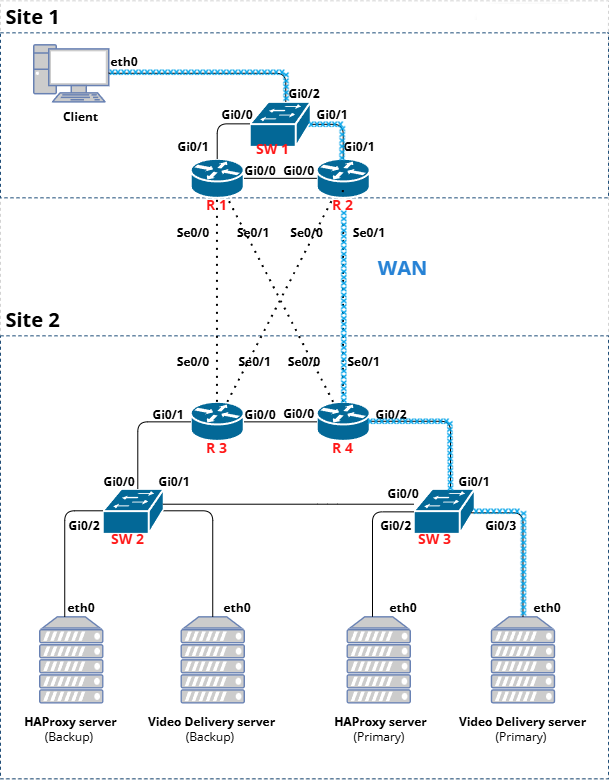}
\label{fig:initial_topology_case_1}
}
\subfigure[Designed topology (Scenario~2).]{
\includegraphics[width=0.3\textwidth]{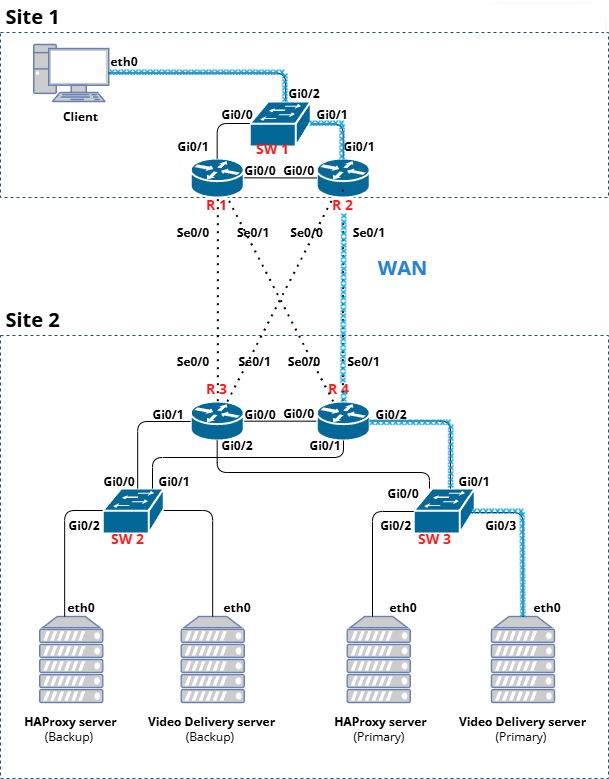}
\label{fig:initial_topology_case_2}
}
\subfigure[LLM-generated topology via prompt]{
\includegraphics[width=0.31\textwidth,]{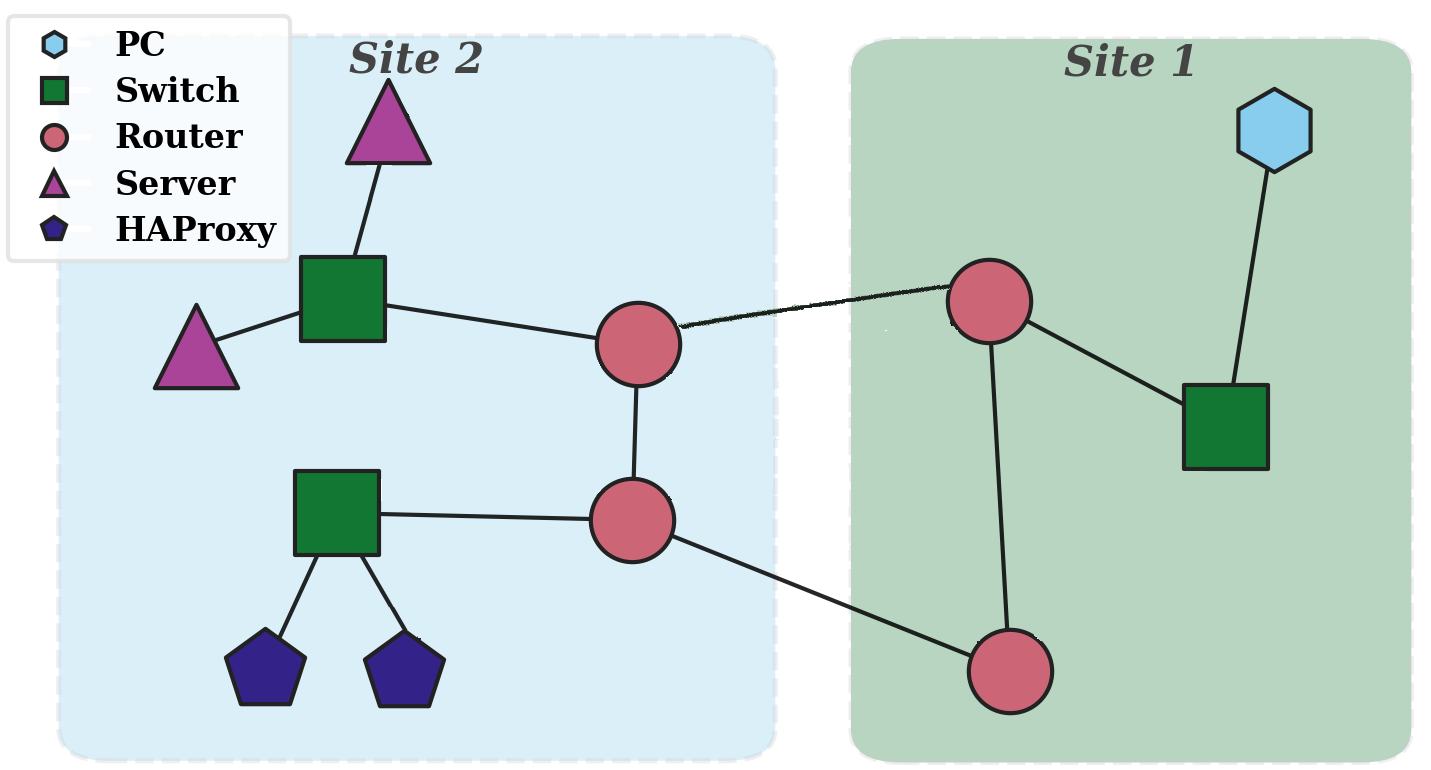}
\label{fig:llm_topology_simple}
}
\caption{Manually designed topologies vs. LLM-generated network topology from a minimal prompt }
\label{fig:initial_topology}
\end{figure*}

\section{Related Work}
\label{sec:related_work}
Recent LLM-based network automation approaches primarily focus on translating natural language intents into device-level configurations, typically assuming a predefined network topology. These methods therefore operate at the configuration layer and do not explicitly address topology synthesis. Network resilience has been extensively studied in classical works such as Hutchison and Sterbenz \cite{hutchison2018architecture}, which define architectural principles for self-configuring and self-healing systems to maintain service continuity under failures. However, these approaches rely on rule-based automation frameworks, which limit their scalability and adaptability in complex network environments. 

Benchmarks such as NetConfEval \cite{netconfeval} and system-level approaches such as S-Witch \cite{jeong2024switch} both illustrate a configuration-centric paradigm in LLM-based network automation, where models operate over predefined network topologies. In this setting, LLMs are primarily used to generate operational commands rather than synthesizing network structures or reasoning about graph-level constraints.
In contrast, our work focuses on the synthesis of structured network topologies under explicit constraints, enabling a complete pipeline from high-level intent to deployable network design. 

Finally, our work relates to studies on the structural generative capabilities of LLMs. Yao et al. \cite{yao2024exploring} investigate graph generation using LLMs and show that models such as GPT-4 can generate structures that satisfy basic topological constraints under controlled prompting. However, their evaluation is limited to abstract and homogeneous graph representations and does not consider the heterogeneity and operational constraints inherent in real-world network infrastructures.
In practical networking scenarios, network topologies are composed of heterogeneous devices such as routers, switches, and servers, each with distinct roles and interface constraints. These architectural properties, together with geographical and administrative constraints, introduce structural requirements that are not fully captured by generic graph generation approaches.




\section{Resilient Topology Generation}
\label{sec:res_gen_methodology}
Failures in networked systems may affect both network devices (e.g., routers, switches, and servers) and communication links (e.g., interface or cable failures), leading to partial or complete service disruption. Ensuring service continuity requires the design of resilient network topologies that incorporate redundancy at both node and link levels. To capture resilience from an operational perspective, we define network robustness in terms of service delivery continuity under failure conditions. Redundancy at the service level can be performed through several paths in the network, but also using multiple service endpoints with primary and backup servers. In this context, we consider two complementary resilience objectives:
\begin{itemize}
    \item \textbf{Server Connectivity ($\textbf{SC}$):} ensuring that each client remains connected to its designated primary server, even in the presence of  failures.
    \item \textbf{Content Connectivity ($\textbf{CC}$):} ensuring that each client is connected to at least one available server (primary or backup) capable of delivering the required service despite failures.
\end{itemize}

\subsection{Problem Formulation}
To illustrate the topology synthesis problem addressed in this work, we consider a user-driven scenario in which an operator aims to design a network distributed across two sites to support a resilient video streaming service. Site~1 hosts a client, a switch, and two routers. Site~2 contains two routers, two switches, two HAProxy load balancers, and two video streaming servers, including primary and backup instances. The service to be delivered by this infrastructure must remain available under network failures (see Figure~\ref{fig:initial_topology}).

Given this specification, the goal is to automatically generate a network topology by extracting all required nodes from the user intent, identifying the role of each device, and determining its corresponding network layer (e.g., access, core layer).
A key requirement is that all connections between devices must be type-compatible (e.g., Ethernet interfaces must be connected to Ethernet interfaces) and explicitly defined in a bidirectional manner, meaning that each link records the exact interfaces used at both endpoints. In addition, the topology generation process must carefully select the set of connections to avoid undesirable loops, which may increase configuration complexity and degrade performance.

From this use case, we define two distinct variations, depending on different hardware configurations: one where all four routers have two serial interfaces and two GigabitEthernet interfaces, and another where they also have an additional GigabitEthernet interface. This variation leads to two distinct resilient topology designs adapted to the available resources (see Figures~\ref{fig:initial_topology_case_1} and~\ref{fig:initial_topology_case_2}). Notably, the second scenario allows each switch to be connected to multiple routers, but this comes with limitations on  inter-switch connectivity and redundant links to prevent routing loops.  Although other topologies could have ensured redundancy, we selected these two designs as they best meet the SC and CC objectives within the given constraints.

Figure~\ref{fig:llm_topology_simple} shows a LLM-generated topology for the \nth{2} scenario. While this topology captures the main network entities, it fails to meet resilience principles. In Site~1, the access switch is connected to a single router, creating a single point of failure. In Site~2, the separation between HAProxy instances and streaming servers prevents proper end-to-end redundancy, since a single switch or switch-to-router link failure may disrupt video delivery. This highlights that simply ensuring basic connectivity, like in a simple graph generation task, is not enough; functional roles and redundancy must also be considered. The topology generation requires joint reasoning over device roles, interface compatibility, structural constraints, and resilience objectives. In this context, we aim to evaluate to what extent LLMs can correctly interpret user requirements and output topology designs that are consistent with such constraint-driven specifications.


\subsection{Topology Model with Interface Constraints}
We model a network topology as a graph \( \mathcal{G} = (\mathcal{V}, \mathcal{E}) \), comprises vertices (devices) \( \mathcal{V} \) and edges (links) \( \mathcal{E} \). Each device \( v_i \in \mathcal{V} \) may have multiple diverse network interfaces, denoted by \( \mathrm{int}_x^{v_i} \), such as a router with Gigabit Ethernet and serial ports.

Classical graph representations model edges as either directed or bidirectional links between homogeneous network nodes. However, such abstractions fall short in representing real-world network configurations involving heterogeneous devices equipped with diverse interfaces. For instance, a link may correspond to a communication channel between a router and a switch (or a server) via connected Gigabit Ethernet interfaces using an Ethernet crossover cable. Alternatively, it may represent a point-to-point communication link between two routers established through a serial cable.
We therefore extend this model by proposing an \textbf{interface-aware formulation} and defining each link $e \in \mathcal{E}$ connecting two end devices $v_i, v_j \in \mathcal{V}$ through interfaces $x$ and $y$ respectively as:
$e = (v_i, \text{int}^{v_i}_x, v_j, \text{int}^{v_j}_y)$.  
This interface-aware formulation enables the enforcement of the following structural constraints:
\begin{itemize}
\item \textbf{Interface exclusivity:} each interface participates in at most one connection.
\item \textbf{Reciprocal consistency:} a link between $(v_i, \text{int}^{v_i}_x)$ and $(v_j, \text{int}^{v_j}_y)$ must be consistently represented at both endpoints.
\item \textbf{No duplicate links:} redundant or conflicting connections are disallowed.
\end{itemize}
Each node in the network topology has a unique identifier, ensuring that it can be consistently referenced across the entire system. To maintain interpretability while respecting LLM context limitations, each node is named in three parts: device type (e.g., router, server), network layer (core, distribution, access, endpoint) and physical location (site).


\subsection{Input Representation: High-Level Intent}

The input consists of high-level natural language descriptions of network requirements written in natural language. A key challenge is the lack of publicly accessible datasets mapping user intent to resilient network topologies.
The input typically includes: multiple geographical sites, device types and their associated interface capabilities, and the types of connections allowed between devices within the same location and between locations.

A representative example is illustrated below:

\begin{tcolorbox}[colback=blue!5!white, colframe=blue!75!black, sharp corners]
\sixptsize
The network consists of two sites: Site 1 and Site 2.  
\textbf{Site 1:} 2 core routers with GigabitEthernet interfaces (Gi0/0, Gi0/1, Gi0/2) and Serial interfaces (Se0/0, Se0/1), 1 access switch, 1 PC.  
\textbf{Site 2:} 2 core routers, 2 access switches, 2 servers.  
\textbf{Connection constraints:} Routers within a site are connected via GigabitEthernet, while inter-site links use Serial interfaces.
\end{tcolorbox}


\subsection{Output Representation}

The generated topology is represented as a structured JSON object capturing the full network design, including devices, links, and interface-level mappings.
The detailed schema of the output format is illustrated below:

\begin{lstlisting}[language=json,basicstyle=\ttfamily\sixptsize,breaklines=true]
{
  "network_topology": {
    "devices": [
      {
        "device_name": "R1",
        "device_type": "Core Router",
        "network_layer": "Core",
        "location_site": "Site 1",
        "device_interfaces": {
          "used_interfaces": [
            {
              "interface_name": "Gi0/0",
              "interface_type": "GigabitEthernet",
              "peer_connection": {
                "peer_device": "R2",
                "peer_interface": "Gi0/0",
                "peer_interface_type": "GigabitEthernet"} }, ... ],
          "unused_interfaces": [] } }, ...  ] }
}
\end{lstlisting}

\begin{figure}[t]
\centering
\includegraphics[width=\columnwidth, height=4cm, keepaspectratio]{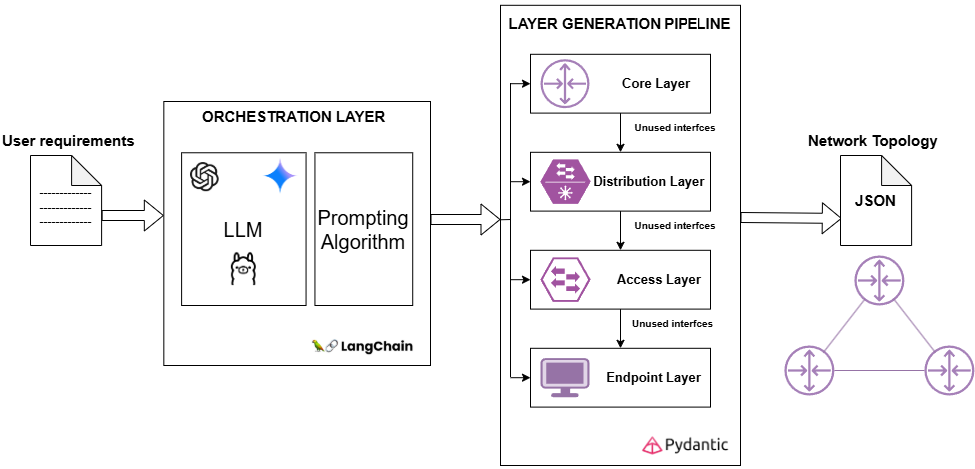}
\caption{ResiNet-LLM: Resilient Topology Generation Framework}
\label{fig:resinet_archi}
\end{figure}

\subsection{ResiNet-LLM: Resilient Topology Generation Framework and Prompt Engineering Strategies}

To address interface tracking complexity and LLM context limitations, we adopt a layered topology generation strategy inspired by hierarchical network design principles. The topology is constructed sequentially across four layers: core, distribution (optional in collapsed-core architectures), access, and endpoint.
At each stage, the LLM is provided only with previously generated nodes and their remaining unused interfaces, ensuring consistent naming, interface validity, and structural coherence.

The overall generation process is implemented in a modular system orchestrated coined as \textbf{ResiNet-LLM}. Built on \textbf{LangChain}, it enforces sequential layer construction, interface propagation, and structural consistency, as illustrated in Figure~\ref{fig:resinet_archi}.
We rely on prompt engineering instead of fine-tuning, given the absence of a dedicated labeled dataset. To ensure structural validity, all generated outputs are validated using \textbf{Pydantic}; when schema violations occur, a feedback mechanism—referred to as \textbf{Pydantic Error Injection}—feeds the error message and previous output back to the model to trigger a corrective regeneration attempt.


To address the multi-constraint nature of resilient topology design, we propose a structured prompting framework with complementary components, each targeting a specific aspect of topology synthesis.
 

\textbf{Role Prompting \& Structured Reasoning:}
The model is first assigned the role of a \textit{Senior Network Architect} to enforce domain-specific constraints. Based on this role, structured reasoning follows a three-stage internal process. In the first stage, the model generates three candidate topologies satisfying all user requirements, while enforcing the following constraints: no isolated nodes, minimization of single points of failure, maximization of redundant paths between sites and key devices, and maximization of path diversity. In the second stage, the model evaluates each candidate against these structural criteria. The selection process follows a strict priority hierarchy: (i) fewer single points of failure, (ii) higher redundant paths, and (iii) higher path diversity. In the last stage, the selected topology is validated through simulated single link and single device failures. If any constraint is violated, the process is restarted from the first stage.



\textbf{Modular Pipeline and State Propagation:}
Each network layer (Core, Distribution, Access, Endpoint) is generated sequentially in a hierarchical manner. We define this mechanism as a \textbf{Stateful Layer-wise Generation Process}. Node identities, device types, and available interface information from the previous layer are propagated to constrain connectivity in the current layer, enforcing controlled dependencies and reducing error propagation across the topology construction pipeline. This mechanism ensures consistency of device naming across layers, as previously generated node identifiers are reused rather than re-invented by the LLM.


\textbf{Structured Output:} We ensure that generated JSON output adheres to a valid schema and structural rules by explicitly defining its format. Specifically, we impose a \textbf{Link Rule}, which enforces bidirectional node connections, ensuring that every link is consistently represented at both endpoints.




\section{Automated Topology Generation Evaluation Framework}
\label{sec:res_gen_evaluation}

\subsection{Node F1 Score}
To evaluate topology reconstruction accuracy, we define a node-level mapping-based F1 metric. Direct node-set comparison is challenging due to naming variations, duplicated entities, or omission of semantically equivalent entities generated by the LLM. Let the reference and generated topologies be \( \mathcal{G}(\mathcal{V}, \mathcal{E}) \) and \( \mathcal{G}_g(\mathcal{V}_g, \mathcal{E}_g) \), respectively. Each node \(v \in \mathcal{V} \cup \mathcal{V}_g\) is described by three structural attributes \( (l_v, s_v, t_v) \), encoding its layer, site, and device type, respectively. We formulate node mapping between \( \mathcal{V} \) and \( \mathcal{V}_g \) as a constrained maximum-weight bipartite matching problem:
\(
M_{\mathcal{V}} \subseteq \{(v,v_g) : v \in \mathcal{V}, v_g \in \mathcal{V}_g, l_v=l_{v_g}, s_v=s_{v_g}, t_v=t_{v_g}\}.
\)
The candidate matching space is restricted to node pairs with compatible structural attributes. Based on attribute alignment, a similarity matrix is constructed, and the Hungarian algorithm \cite{Hungarian55} is applied to compute the maximum matching \(M_{\mathcal{V}}\), yielding a one-to-one node mapping. A reference node \(v \in \mathcal{V}\) is considered correctly reconstructed if it is mapped to a semantically consistent generated node under \(M_{\mathcal{V}}\). Mapped nodes are counted as True Positives (TP), unmatched generated nodes as False Positives (FP), and unmatched reference nodes as False Negatives (FN). The node-level F1 score is computed as:
\[
F1_{nodes} = \frac{2|TP|}{2|TP|+|FP|+|FN|}.
\]
This metric measures node recovery accuracy while being robust to naming variations and preserving semantic consistency in hierarchical topologies.

\subsection{Edge F1 Score}

The edge-level evaluation measures whether the generated topology preserves the connectivity structure of the reference graph. Compared with node-level matching, edge validation requires considering endpoint correspondence, interface compatibility, and bidirectional connectivity constraints. Given the maximum matching  \(M_{\mathcal{V}}\) (node mapping), edge correspondence is progressively refined. Edges connected to unmatched reference nodes are treated as missing edges (FN), while edges involving unmatched generated nodes are considered spurious edges (FP). For the remaining candidate edges, validity is determined by endpoint alignment under \(M_{\mathcal{V}}\), bidirectional connectivity, and interface compatibility. Edges violating these constraints are classified as incorrect.

To handle hierarchical topologies where functionally equivalent nodes may be permuted, we further refine node correspondence among nodes sharing the same layer, site, and device type. Each node is represented by a connectivity signature, defined as a vector counting neighboring nodes by device type. Nodes with similar signatures are aligned through the refinement of \(M_{\mathcal{V}}\), without introducing an additional edge-level optimization. This ensures that structurally equivalent configurations remain comparable and that edge evaluation is independent of arbitrary node ordering.

Based on the resulting correspondence, correctly mapped edges are counted as TPs, missing reference edges as FNs, and spurious or invalid generated edges as FPs. From these counts, the same formulation as the node-level metric is used to compute the edge-level F1 score. This score evaluates connectivity reconstruction accuracy while accounting for node mapping, interface constraints, and global structural consistency.

\subsection{Resilience Quantification Framework}
To evaluate the robustness of LLM-generated topologies under failure conditions, we assess their functional resilience and connectivity preservation.

\textbf{Criticality-Aware Stress Testing:}  
We assume that a client \(c\) accesses video content from the primary or backup delivery server (\(S_p\) or \(S_b\)) through the shortest path \(SP(c,S)\), where \(S \in \{S_p,S_b\}\). The corresponding sets of nodes and edges in  \(SP(c,S)\) are given as: $\mathcal{V}_{SP}=\{v:v\in SP(c,S)\}$, $\mathcal{E}_{SP}=\{e:e\in SP(c,S)\}$. Single-point-of-failure scenarios are simulated by removing nodes or links along \(SP(c,S)\) and verifying whether connectivity objectives are preserved. Since some components (e.g., client endpoints and access switches) cannot be replicated, they are manually identified as critical nodes \(\mathcal{V}_{crit}\). The resilience evaluation sets are then defined as:
$\mathcal{V}_{eval}=\mathcal{V}_{SP}\setminus\mathcal{V}_{crit}$, $\mathcal{E}_{eval}=\mathcal{E}_{SP}\setminus\mathcal{E}(\mathcal{V}_{crit})$, where \(\mathcal{E}(\mathcal{V}_{crit})\) denotes edges connecting two critical nodes.

\textbf{Resilience Metrics: }Resilience is evaluated using shortest-path analysis under two connectivity objectives: Server Connectivity (SC) and Content Connectivity (CC). For a failed network element \(ne\) (a node \(v\) or an edge \(e\)) and a connectivity objective \(obj \in \{\mathrm{SC},\mathrm{CC}\}\), we define \(f(ne,\mathrm{obj})=1\) if an alternative path from \(c\) to \(S\) preserves the objective after the failure, and \(0\) otherwise. The \textbf{node and link resilience ratios}, denoted by \(R_{\mathcal{V},\mathrm{obj}}\) and \(R_{\mathcal{E},\mathrm{obj}}\), respectively, are then computed as:
\[
R_{\mathcal{V},\mathrm{obj}}=
\frac{1}{|\mathcal{V}_{eval}|}
\sum_{v\in\mathcal{V}_{eval}} f(v,\mathrm{obj}),
\]

\[
R_{\mathcal{E},\mathrm{obj}}=
\frac{1}{|\mathcal{E}_{eval}|}
\sum_{e\in \mathcal{E}_{eval}} f(e,\mathrm{obj}).
\]
If the generated topology is initially disconnected, the corresponding resilience ratios are set to zero. This separation of node and link failures provides a granular quantitative evaluation of topology resilience under single-failure scenarios. For illustration, consider the topology in Figure~\ref{fig:initial_topology_case_1}, where a client at Site 1 accesses the primary server at Site 2 through the shortest path  which is highlighted by the bold blue line. In this case, \(\mathcal{V}_{crit}=\{\texttt{Client},\texttt{SW1}\}\), and only the link between \texttt{Client} and \texttt{SW1} is excluded from edge evaluation. The resulting resilience ratios are
\resizebox{\linewidth}{!}{%
$R_{\mathcal{V},\text{SC}}=2/4$, $R_{\mathcal{V},\text{CC}}=4/4$, $R_{\mathcal{E},\text{SC}}=3/4$, $R_{\mathcal{E},\text{CC}}=4/4$%
}.
\begin{table}[t]
\centering
\caption{Description of the 4 Test Scenarios}
\label{tab:f1_llms_onecol}
\resizebox{\columnwidth}{!}{%
\begin{tabular}{lrrll}
\toprule
\textbf{Scenario} & \textbf{\# Nodes} & \textbf{\# Links} & \textbf{\# Network Devices} & \textbf{\# Endpoint Devices} \\
\midrule
1 & 12   & 16  & 4\,R, 3\,AS  & 4\,S, 1\,PC \\
2 & 12   & 17  & 4\,R, 3\,AS  & 4\,S, 1\,PC \\
3 & 32   & 42  & 8\,R, 4\,AS, 4\,WAP & 8\,P, 8\,PC \\
4 & 173  & 178 & 4\,R, 5\,AS & 4\,S, 80\,IP, 80\,PC \\
\bottomrule
\end{tabular}%
}

\begin{tablenotes}
\footnotesize
\item 
\textbf{R}: Router, \textbf{AS}: Access Switch, 
\textbf{WAP}: Wireless Access Point, \textbf{S}: Server, \textbf{PC}: Personal Computer, \textbf{IP}: IP Phone, \textbf{P}: Printer.
\end{tablenotes}
\end{table}

\begin{table*}[t]
\centering
\caption{Connectivity Resilience Across Prompt Engineering Techniques (Scenario 2)}
\label{tab:f1_llms_two_col_empty}
\setlength{\tabcolsep}{2pt} 
\renewcommand{\arraystretch}{1.1}

\resizebox{\textwidth}{!}{%
\begin{tabular}{c|c|r@{$\pm$}lr@{$\pm$}lr@{$\pm$}l|r@{$\pm$}lr@{$\pm$}lr@{$\pm$}l|r@{$\pm$}lr@{$\pm$}lr@{$\pm$}l}
\toprule

& & \multicolumn{6}{c|}{\textbf{GPT-4o}} & \multicolumn{6}{c|}{\textbf{Mistral-small-24B}} & \multicolumn{6}{c}{\textbf{Qwen3-32B}} \\

\cmidrule(lr){3-8}\cmidrule(lr){9-14}\cmidrule(l){15-20}

\textbf{Task} & \textbf{Ablation}
& \multicolumn{2}{c}{\textbf{F1}} & \multicolumn{2}{c}{\textbf{SC}} & \multicolumn{2}{c|}{\textbf{CC}}
& \multicolumn{2}{c}{\textbf{F1}} & \multicolumn{2}{c}{\textbf{SC}} & \multicolumn{2}{c|}{\textbf{CC}}
& \multicolumn{2}{c}{\textbf{F1}} & \multicolumn{2}{c}{\textbf{SC}} & \multicolumn{2}{c}{\textbf{CC}} \\

\midrule

\multirow{6}{*}{\rotatebox{90}{\textbf{Node}}}
& Baseline Prompt
& 0.58 & 0.00 & 28\% & 4\%  & 70\% & 24\%
& 0.59 & 0.02 & 3\% & 11\% & 23\%    & 31\%
& 0.60 & 0.06 & 21\% & 7\%  & 44\%    & 12\% \\

& Full Prompt
& \textbf{1.00} & \textbf{0.00} & \textbf{35\%} & \textbf{12\%} & \textbf{85\%}    & \textbf{12\%}
& \textbf{0.98} & \textbf{0.07} & \textbf{17\%} & \textbf{21\%} & \textbf{45\%}    & \textbf{43\%}
& 0.92 & 0.10 & 32\% & 21\% & 52\%    & 32\% \\

& Without Structured reasoning
& \textbf{1.00} & \textbf{0.00} & 25\%  & 0\%  & 65\%    & 20\%
& 0.97 & 0.08 & 2\%  & 11\% & 5\%     & 22\%
& 0.88 & 0.11 & 4\%  & 16\% & 5\%     & 22\% \\

& Without Stateful Layering
& \textbf{1.00} & \textbf{0.00} & 20\% & 10\% & 80\%    & 40\%
& 0.40 & 0.49 & 10\% & 19\% & 17\%    & 32\%
& \textbf{1.00} & \textbf{0.00} & \textbf{42\%} & \textbf{11\%} & \textbf{67\%}    & \textbf{11\%} \\


& Without Pydantic Error Injection
& \textbf{1.00} & \textbf{0.00} & \textbf{35\%} & \textbf{12\%} & \textbf{85\%}    & \textbf{12\%}
& 0.53 & 0.11 & 0\%  & 0\%  & 2\%     & 11\%
& 0.50 & 0.00 & 0\%  & 0\%  & 0\%     & 0\% \\

\midrule
\multirow{6}{*}{\rotatebox{90}{\textbf{Edge}}}
& Baseline Prompt
& 0.47 & 0.00 & 57\% & 8\%  & \textbf{85\%}    & \textbf{12\%}
& 0.40 & 0.06 & 14\% & 20\% & 28\%    & 36\%
& 0.48 & 0.05 & 44\% & 12\% & 67\%    & 17\% \\

& Full Prompt
& \textbf{0.89} & \textbf{0.05} & \textbf{60\%} & \textbf{12\%} & \textbf{85\%}    & \textbf{12\%}
& \textbf{0.72} & \textbf{0.18} & \textbf{39\%} & \textbf{36\%} & \textbf{52\%}    & \textbf{48\%}
& 0.74 & 0.17 & 55\% & 32\% & 74\%    & 43\% \\

& Without Structured reasoning
& 0.83 & 0.05 & 25\% & 0\%  & 70\%    & 10\%
& 0.63 & 0.18 & 4\%  & 16\% & 5\%     & 22\%
& 0.71 & 0.21 & 4\%  & 16\% & 5\%     & 22\% \\

& Without Stateful Layering
& 0.88 & 0.06 & 40\% & 20\% & 80\%    & 40\%
& 0.30 & 0.38 & 16\% & 28\% & 21\%    & 37\%
& \textbf{0.80} & \textbf{0.06} & \textbf{67\%} & \textbf{11\%} & \textbf{92\%}    & \textbf{11\%} \\


& Without Pydantic Error Injection
& \textbf{0.89} & \textbf{0.05} & \textbf{60\%} & \textbf{12\%} & \textbf{85\%}    & \textbf{12\%}
& 0.33 & 0.10 & 2\%  & 11\% & 4\%     & 16\% 
& 0.49 & 0.06 & 0\%  & 0\%  & 0\%     & 0\% \\

\bottomrule
\end{tabular}%
}

\end{table*}

\section{Experimental Results and Analysis}
\label{sec:results topology}

We evaluate our approach using five LLMs: Gemini 2.5 Flash (Gemini), GPT-4o (GPT), Mistral-Small-24B (Mistral), Qwen3-32B (Qwen), and DeepSeek-R1-32B (DeepSeek), covering both proprietary and open models with diverse reasoning capabilities. Gemini and GPT represent state-of-the-art proprietary systems accessed via native interfaces or APIs, while Mistral, Qwen, and DeepSeek are competitive open models executed locally using LangChain and Ollama on a single NVIDIA RTX 4500 Ada GPU (24 GB VRAM). We consider four test scenarios summarized in Table~\ref{tab:f1_llms_onecol}. The temperature is set to 0.3 to reduce randomness while maintaining diversity. Context windows are set to 8,192 tokens for Sce. 1--2 and 16,384 tokens for Sce. 3--4 to support increasing topology complexity.


For evaluation, we first focus on Sce. 2, whose reference topology is shown in Figure~\ref{fig:initial_topology_case_2}, to assess the effectiveness of the proposed ResiNet-LLM framework for realistic network topology generation. This includes an ablation study on prompt engineering strategies, followed by quantitative and qualitative evaluations of the full prompt technique. We further validate the framework using two additional LLMs and enterprise-inspired scenarios with varying network sizes and architectures. Each scenario is paired with a reference topology satisfying the same requirements, enabling consistent comparison via a unified JSON file. The evaluation scripts are available on our Hugging Face repository\footnote{\url{https://huggingface.co/ResiNetResearchGroup/ResiNet-LLM-topology}}.

\subsection{Ablation Study of the Prompt Engineering}
At first, we evaluate the contribution of each component of the proposed ResiNet-LLM framework through an ablation study on Scenario~2, using a fixed scenario to ensure that all performance variations are attributable to the prompting strategy. Starting from a baseline prompt, we evaluate the full configuration and progressively remove individual components to quantify their impact on topology generation.
This study evaluates three models---Mistral, Qwen, and GPT---with the results reported in Table~\ref{tab:f1_llms_two_col_empty} as the mean and standard deviation across multiple runs.

\textbf{Baseline Prompt: } When tested with the baseline prompt---restricted to role, scenario and format output descriptions---all models consistently produce structural and semantic errors (see Figure~\ref{fig:llm_topology_simple}). Mistral exhibits strong connectivity-related failures, including graph disconnections, node count mismatches, and bidirectionality violations, resulting in low CS and CC scores. These structural issues directly degrade its node and edge F1 scores. GPT achieves higher CC but lower CS, indicating structurally consistent but not fully resilient topologies. Qwen shows relatively stronger CC at the edge level, while its CS remains weaker, reflecting partial structural correctness with limited robustness.
\begin{figure}[t]
\centering
\subfigure[Full Prompting.]{
\includegraphics[width=0.45\columnwidth]{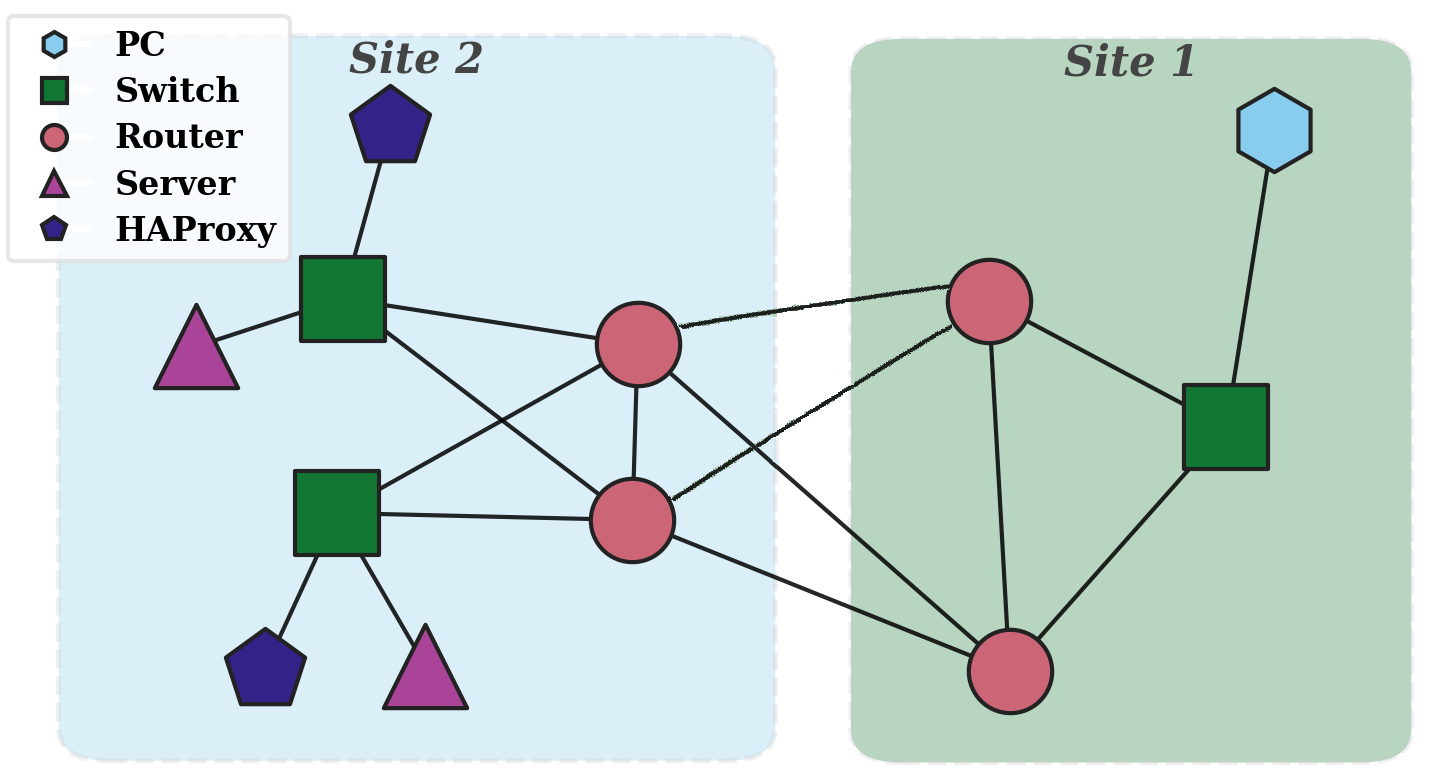}
\label{fig:mistral_full_prompt_test_4}
}
\subfigure[w/o Structured Reasoning.]{
\hfill
\includegraphics[width=0.45\columnwidth]
{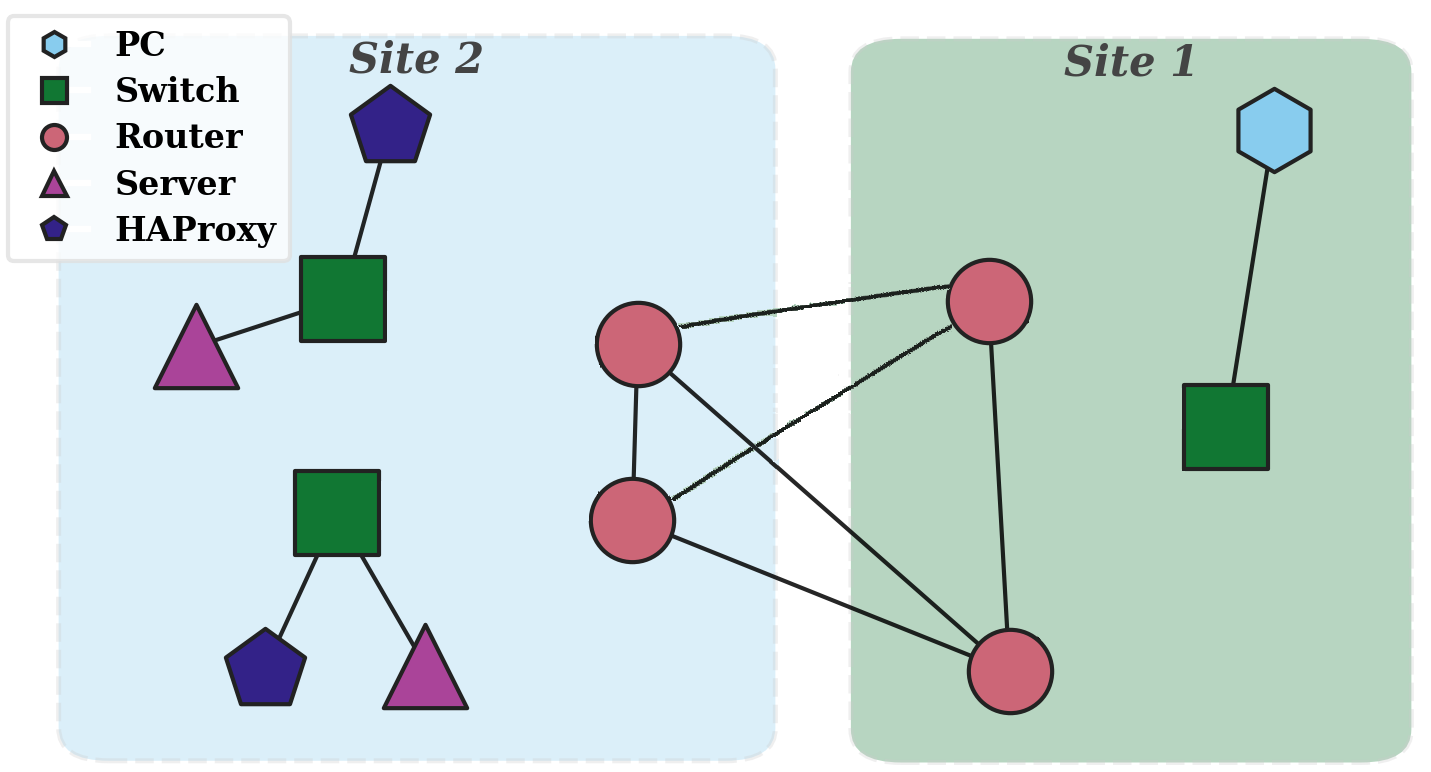}
\hfill
\label{fig:mistral_sans_TH_test_11}
}
\label{fig:imistral_topologies}
\caption{Examples of topologies generated by Mistral}
\end{figure}

\textbf{Full Prompt: } The full prompt leads to consistent improvements across all models. For GPT, node and edge F1 scores increase due to the elimination of layer assignment errors, while CC remains stable and CS improves thanks to better redundancy modeling and correct placement of end components. Mistral shows a significant increase in node F1 scores, although performance remains sub-optimal due to occasional structural inconsistencies induced by JSON generation errors. Edge F1 also improves, alongside gains in both CS and CC. Qwen exhibits similar improvements in node F1, despite being more affected by JSON-related inconsistencies. At the edge level, it outperforms Mistral, with notable increases in both CS and CC, even when the node F1 score remains lower. To illustrate these improvements qualitatively, Figure~\ref{fig:mistral_full_prompt_test_4} presents an example topology generated by all LLMs under the full prompt, which perfectly reproduces the reference structure given in Figure~\ref{fig:initial_topology_case_2}.

\textbf{Without Structured Reasoning: } Removing the Structured Reasoning component reduces performance across all evaluated models. For GPT, despite maintaining a perfect node F1 score, the absence of structured reasoning reduces CS and CC by limiting the preservation of alternative paths. For Mistral, the performance decrease is mainly attributed to the generation of non-fully connected graphs, resulting in lower edge F1 and a slight reduction in node F1, which jointly affect CS and CC. This issue is illustrated by an example topology generated by Mistral under this configuration, shown in Figure~\ref{fig:mistral_sans_TH_test_11}. For Qwen, a more significant degradation is observed due to erroneous JSON-based generation, leading to missing nodes.

\textbf{Without Stateful Layering: } Removing the stateful layering mechanism yields model-dependent effects. For GPT, node F1 remains perfect while edge F1 drops slightly; this causes a minor drop in CC but a stronger decrease in CS, showing that even a single non-redundant link can significantly impact structural robustness. For Mistral, node F1 decreases significantly, even when increasing the context window, as the model fails to generate topologies across multiple inferences, producing empty graphs and dragging down both CS and CC. In contrast, Qwen's F1, CS, and CC improve due to better node generation and increased path redundancy. Notably, while the node count in scenario~2 is smaller than those in scenario 3 and 4, attempting to process all layers in a single generation drastically degrades performance for larger topologies, particularly for Mistral and Qwen.


\textbf{Without Pydantic Error Injection: }As GPT consistently produces valid JSON outputs, its F1 scores, CS and CC remain unchanged compared to full prompt. In contrast, Mistral and Qwen exhibit a drop in node F1, which propagates to edge F1 and results in reduced connectivity.

\begin{table}[t]
\centering
\caption{Topology Generation Performance and Validation Across LLMs in Scenario 2.}
\label{tab:resilience_final}
\resizebox{\columnwidth}{!}{%
\begin{tabular}{l|r@{$\pm$}lr@{$\pm$}lr@{$\pm$}lr@{$\pm$}lr@{$\pm$}l}
\toprule
\textbf{Metrics} 
& \multicolumn{2}{c|}{\textbf{\makecell{Gemini\\2.5 flash}}} 
& \multicolumn{2}{c|}{\textbf{GPT-4o}}  
& \multicolumn{2}{c|}{\textbf{\makecell{Mistral-small-24B}}} 
& \multicolumn{2}{c|}{\textbf{Qwen3-32b}} 
& \multicolumn{2}{c}{\textbf{\makecell{DeepSeek-R1-32b}}} \\
\midrule
Valid                & \multicolumn{2}{c|}{$\textbf{5/5}$} & \multicolumn{2}{c|}{$2/5$}  & \multicolumn{2}{c|}{$2/20$} & \multicolumn{2}{c|}{$2/20$} & \multicolumn{2}{c}{$3/20$} \\
Partial redundancy   & \multicolumn{2}{c|}{$0/5$}           & \multicolumn{2}{c|}{$3/5$}  & \multicolumn{2}{c|}{$9/20$} & \multicolumn{2}{c|}{$14/20$} & \multicolumn{2}{c}{$0/20$}  \\
Connectivity Failure & \multicolumn{2}{c|}{$0/5$}           & \multicolumn{2}{c|}{$0/5$}  & \multicolumn{2}{c|}{$8/20$} & \multicolumn{2}{c|}{$1/20$}  & \multicolumn{2}{c}{$14/20$} \\
Incomplete           & \multicolumn{2}{c|}{$0/5$}           & \multicolumn{2}{c|}{$0/5$}  & \multicolumn{2}{c|}{$1/20$} & \multicolumn{2}{c|}{$3/20$}  & \multicolumn{2}{c}{$3/20$} \\
\midrule
\textbf{SC (Nodes)} & \textbf{50\%} & \textbf{0\%} & 35\% & 12\% & \ \ \ \ 17\% & 21\% & 32\% & 21\% & \ \ \ 10\% & 20\% \\
\textbf{CC (Nodes)} & \textbf{100\%} & \textbf{0\%} & 85\% & 12\% & 45\% & 43\% & 52\% & 32\% & 20\% & 40\% \\
\textbf{SC (Edges)} & \textbf{75\%} & \textbf{0\%} & 60\% & 12\% & 39\% & 36\% & 55\% & 32\% & 15\% & 30\% \\
\textbf{CC (Edges)} & \textbf{100\%} & \textbf{0\%} & 85\% & 12\% & 52\% & 48\% & 73\% & 43\% & 20\% & 40\%\\
\hline
\textbf{Generation Time (s)} & \multicolumn{2}{c|}{$\sim$20} & \multicolumn{2}{c|}{$\sim$21} & \multicolumn{2}{c|}{$\sim$114} & \multicolumn{2}{c|}{$\sim$502} & \multicolumn{2}{c}{$\sim$317} \\
\bottomrule
\end{tabular}%
}
\end{table}

\subsection{Quantitative and Qualitative Evaluations}
We next extend the full-prompt analysis to additional models (Gemini and DeepSeek) on Scenario 2, enabling a broader comparison of topology generation behavior across LLMs. We report quantitative SC/CC metrics for both nodes and edges, complemented by a qualitative classification of generated topologies to characterize structural failure modes. Specifically, outputs are categorized as \textit{Valid}, \textit{Partial redundancy}, \textit{Connectivity failure}, or \textit{Incomplete}, corresponding respectively to fully correct graphs, structurally complete but non-redundant graphs, disconnected graphs, and graphs with missing nodes.
Table~\ref{tab:resilience_final} summarizes both quantitative metrics, qualitative distributions, and generation time. Gemini consistently produces valid topologies, matching reference-level SC and CC, although it occasionally generates extra inter-switch links that do not alter the resulting connectivity structure. GPT shows a high proportion of valid or near-valid outputs. Mistral and DeepSeek are dominated by connectivity failures, while Qwen is mainly characterized by partial redundancy. Overall, connectivity failures are primarily associated with structural violations (e.g., bidirectionality), while partial redundancy reflects complete but less resilient graph constructions. DeepSeek additionally exhibit JSON-related inconsistencies, that cause node-level degradation and incomplete graphs. Generation time varies significantly across models: API-based models (GPT, Gemini) are faster, while local models (Mistral, Qwen, DeepSeek) exhibit higher latency due to inference constraints.






\begin{table}[t]
\centering
\caption{F1 scores for Network Topology Generation (Node vs. Edge) in 4 Scenarios}
\label{tab:f1_node_edge}
\resizebox{\columnwidth}{!}{%
\begin{tabular}{cclllll}
\toprule
\textbf{Task} & \textbf{Scenarios} & \textbf{\makecell{Gemini\\2.5 flash}} & \textbf{GPT-4o} & \textbf{Mistral-small-24B} & \textbf{Qwen3-32B} & \textbf{DeepSeek-R1-32B} \\
\midrule

\multirow{4}{*}{\rotatebox{90}{\textbf{Node}}} 
& Scenario 1 & \textbf{1.00 $\pm$ 0.00} & \textbf{1.00 $\pm$ 0.00}  & \textbf{1.00 $\pm$ 0.00}  & 0.94 $\pm$ 0.07  & 0.97 $\pm$ 0.06 \\

& Scenario 2 & \textbf{1.00 $\pm$ 0.00} & \textbf{1.00 $\pm$ 0.00} & 0.98 $\pm$ 0.07 & 0.92 $\pm$ 0.10 & 0.97 $\pm$ 0.06 \\


& Scenario 3 & \textbf{1.00 $\pm$ 0.00}  & 0.93 $\pm$ 0.00 & 0.91 $\pm$ 0.07 & 0.29 $\pm$ 0.28 & 0.76 $\pm$ 0.14 \\

& Scenario 4 & \textbf{0.22 $\pm$ 0.00}  & \textbf{0.22 $\pm$ 0.00} & 0.10 $\pm$ 0.00 & 0.10 $\pm$ 0.00  & 0.16 $\pm$ 0.00 \\

\midrule 

\multirow{4}{*}{\rotatebox{90}{\textbf{Edge}}} 
& Scenario 1 & \textbf{1.00 $\pm$ 0.00} & 0.95 $\pm$ 0.02 & 0.78 $\pm$ 0.11 & 0.73 $\pm$ 0.09 & 0.67 $\pm$ 0.15 \\
& Scenario 2 & \textbf{0.97 $\pm$ 0.00} & 0.89 $\pm$ 0.05 & 0.72 $\pm$ 0.18 & 0.74 $\pm$ 0.17 & 0.58 $\pm$ 0.15 \\
& Scenario 3 & \textbf{0.88 $\pm$ 0.05} & 0.72 $\pm$ 0.00 & 0.56 $\pm$ 0.25 & 0.11 $\pm$ 0.09 & 0.52 $\pm$ 0.21 \\
& Scenario 4 & \textbf{0.10 $\pm$ 0.00} & \textbf{0.10 $\pm$ 0.00} & 0.01 $\pm$ 0.00 & 0.06 $\pm$ 0.00 & 0.06 $\pm$ 0.00 \\

\bottomrule
\end{tabular}%
}
\end{table}

\subsection{Generalization across four scenarios}

To evaluate the robustness of the proposed full prompt, we conduct experiments across the four scenarios in Table~\ref{tab:f1_llms_onecol}. Since each scenario's structural constraints allow for multiple valid, resilient topologies, we use a single representative reference topology per scenario to ensure consistent evaluation. 
While prior experiments required 20 runs per configuration to capture SC and CC variability in open-weight models, we use 5 runs per scenario here. F1-scores exhibit greater stability, allowing this reduced sample size to minimize computational overhead without compromising reliability.

We report node and edge F1-scores (Table~\ref{tab:f1_node_edge}) as our primary structural metrics. SC and CC are omitted since they are derived from connectivity and offer limited additional discrimination at the scenario level. Scenario~1, structurally similar to Scenario~2 but with fewer router interfaces, yields consistently high performance across all models, particularly at the edge level. This confirms that reducing connectivity complexity improves structural accuracy. Furthermore, F1-scores remain high and decrease only gradually as network size increases up to Scenario~3 (32 nodes), before dropping sharply in Scenario~4. This trend indicates that while the proposed framework performs well for medium-scale networks, its performance degrades in large-scale scenarios, highlighting scalability limitations in complex topologies.




\section{Conclusions}

This paper investigated the use of LLMs for zero-touch resilient network topology generation from high-level natural language requirements. We proposed ResiNet-LLM, a framework combining stateful layer-wise generation, structured reasoning, schema-based validation, and iterative error correction to enable constraint-aware topology synthesis. Experiments in four representative scenarios demonstrate that the proposed approach consistently improves topology generation quality across various models. While proprietary models achieve the highest stability and accuracy, open-weight models also benefit from the framework though they remain prone to structural inconsistencies under complex constraints. The results further show that combining structural and connectivity-based metrics provides a more comprehensive evaluation of generated topologies than structural similarity alone. 


\bibliographystyle{IEEEtran}
\bibliography{ref}

\end{document}